\documentclass[12pt]{iopart}
\usepackage{iopams}
\usepackage{graphicx}

\begin{document}

\title{Vacuum shell in the Schwarzschild-de\,Sitter world}

\author{S V Chernov$^{\dag\ddag}$\ and V I Dokuchaev$^\dag$}

\address{${\dag}$\ Institute for Nuclear Research of the
Russian Academy of Sciences \\
60th October Anniversary Prospect 7a, 117312 Moscow, Russia \\
\ddag\ Moscow Institute of Physics and Technology \\ Dolgoprudny,
141700 Moscow region, Russia} \ead{chernov.s@mail.ru,
dokuchaev@inr.npd.ac.ru}

\begin{abstract}
We construct the classification scheme for all possible evolution
scenarios and find the corresponding global geometries for
dynamics of a thin spherical vacuum shell in the
Schwarzschild-de\,Sitter metric. This configuration is suitable
for the modelling of vacuum bubbles arising during cosmological
phase transitions in the early Universe. The distinctive final
types of evolution from the local point of view of a rather
distant observer are either the unlimited expansion of the shell
or its contraction with a formation of black hole (with a central
singularity) or wormhole (with a baby universe in interior).
\end{abstract}

%\submitto{\JPG}

\maketitle

\section{Introduction}

One of the promising mechanisms of primordial black holes and
wormholes formation is a collapse of vacuum bubbles during phase
transitions in the early Universe
\cite{ZeldovKobzarOcyn,VolKobOcyn,Col,CalCol,ColLuc}. The
supermassive black holes in the centers of galaxies may be also
originated by this mechanism. We will describe the possible types
of dynamical evolution of vacuum bubbles in the
Schwarzschild-de\,Sitter metric by using the thin shell
approximation for the boundary between the true and false vacuum.
The formalism of thin shells in General Relativity was first
developed by W. Israel \cite{Israel}. Later this formalism was
elaborated and adjusted for the case of cosmological vacuum phase
transitions
\cite{BerKuzTkach1,BerKuzTkach2,BerKuzTkach3,BerKuzTkach4}.

The boundary of vacuum bubble divides the
Schwarz\-schild-de\,Sitter space-time into internal and external
regions. In the following these regions are designated by indexes
``$in$'' and ``$out$'' respectively. Our aim is a full
classification of possible evolution scenarios of vacuum bubbles
versus of parameters of the Schwarzschild-de\,Sitter space-time
and initial conditions for a thin shell. It appears that a useful
classification quantity for this problem is a mass parameter $m$
which will be defined below. Depending on this mass parameters the
bubble is either expanding to infinity or contracting with a final
formation of black hole or wormhole
\cite{SatoSasKodMae,SatoSasKodMae1,FrolMarMuk}. This work is a
generalization of the earlier analysis in
\cite{BlGuenGuth,AguMat,Lee-Park,DokCher}, where only some
particular scenarios for this problem were investigated.

In previous works
\cite{BerKuzTkach4,SatoSasKodMae,SatoSasKodMae1,BlGuenGuth,AguMat,AurPalSpal,AurJoh}
were considered only the particular cases of solutions
corresponding to the zero value of our inner mass parameter
$m_{in}=0$. We describe a more general case when $m_{in}\neq0$,
and then the new types of solutions appear. The bubbles are
originated in the phase transition in the early universe and may
contain in principle smaller bubbles inside (see e. g.
\cite{SatoSasKodMae,SatoSasKodMae1}). To model this in a formal
way we include an interior mass parameter $m_{in}$. This parameter
may be considered as a seed black hole.

In general, solutions with $m_{in}\neq0$ are quite the similar
ones to considered in
\cite{BerKuzTkach4,SatoSasKodMae,SatoSasKodMae1,BlGuenGuth,AguMat,AurPalSpal,AurJoh}.
At the same time the addition of inner mass parameter $m_{in}$
results in a complication of the classification scheme for
possible solutions.

In Section~\ref{sec2} the equation of motion for a thin shell in
the Schwarzschild-de\,Sitter metric is analyzed. Basing on this
equation in Section~\ref{sec3} we develop a classification scheme
for possible evolution scenarios and construct also the
Carter-Penrose diagrams for corresponding global geometries. The
concluding remarks are shortly summa\-rized in Section~\ref{sec4}.

\section{Equation of motion}
\label{sec2}

\begin{figure}[t]
\begin{center}
\includegraphics[width=0.8\textwidth]{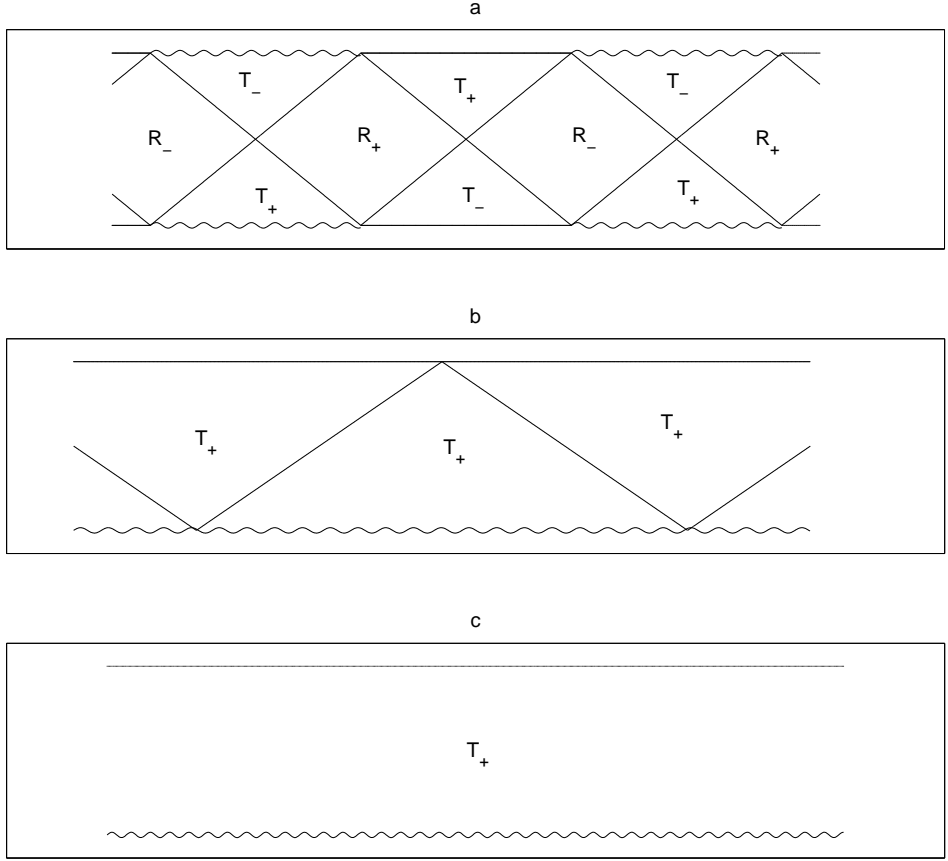}
\end{center}
\caption{\label{figure1} The Carter-Penrose diagrams for the
Schwarzschild-de Sitter metric. In the case (a), when $M<m_{2}$,
there are event horizons. There is only one event horizon in the
case (b), when $M=m_{2}$. In the case (c), when $M>m_{2}$, the
event horizon is absent. Diagrams (b) and (c) are shown for the
case of expansion. The regions $T_{+}$ and $T_{-}$ and also zero
and infinity lines are swapped in the corresponding (nonshown)
diagrams for the case of contraction. The oppositely directed null
lines in these diagrams and in the following ones are not
necessarily perpendicular to each other due to arbitrary
deformation of the coordinate systems.}
\end{figure}

The Schwarzschild-de\,Sitter metric can be written in the form
\begin{equation}
 ds^{2} = \left(1-\frac{2M}{r}-\frac{8\pi}{3}\varepsilon
 r^2\right)dt^2 - \left(1-\frac{2M}{r}-
 \frac{8\pi}{3}\varepsilon r^2\right)^{-1}\!dr^2-r^2d\Omega,
 \label{MetSharDeSit}
\end{equation}
where $M$ is the Schwarzschild mass, $\varepsilon$ is a vacuum
energy density and $d\Omega=d\theta^2+\sin^2\theta\, d\phi^2.$
This metric has the following specific properties. The positive
roots of equation
\begin{equation}
 1-\frac{2M}{r}-\frac{8\pi\varepsilon}{3}\,r^2=0
 \label{horizon}
\end{equation}
define the radii of event horizons in this metric. The number of
positive roots (and so the number of event horizons) depends on
the ratio of $M$ and $1/\sqrt{\varepsilon}$. There are no event
horizons, if $M>m_{2}=1/\sqrt{72\pi\varepsilon}$. There is only
one event horizon, $r_{\rm h1}=1/\sqrt{8\pi\varepsilon}$, if
$M=m_{2}$. In the case $M<m_{2}$ there are two distinctive event
horizons:
\begin{eqnarray}
 r_{\rm h2} &=& 2\sqrt{\frac{p}{3}}\cos\left[\frac{\pi}{3}+
 \frac{1}{3}\arctan\sqrt{\frac{4p^3}{27q^2}-1}\right];  \\
 r_{\rm h3} &=& 2\sqrt{\frac{p}{3}}\cos\left[\frac{\pi}{3}-
 \frac{1}{3}\arctan\sqrt{\frac{4p^3}{27q^2}-1}\right],
\end{eqnarray}
where $p=3/(8\pi\varepsilon)$ and $q=3M/(4\pi\varepsilon)$. It can
be shown that $r_{\rm h2}<r_{\rm h1}<r_{\rm h3}<\sqrt{3}r_{\rm
h1}$.  See in the Fig.~\ref{figure1} the corresponding
Carter-Penrose diagrams for global geometry of the
Schwarzschild-de\,Sitter space-time \cite{BerKuzTkach3,AguMat}.

The basic equation of motion for a thin vacuum shell in the
Schwarzschild-de Sitter space-time, resulting from the matching of
the inner and outer metrics on the shell, can be written in the
following form
\cite{BerKuzTkach1,BerKuzTkach2,BerKuzTkach4,Lee-Park,Chao,AurilDen,Alb-Trod,Gon}:
\begin{equation}
 4\pi\sigma\rho=\sigma_{\rm in}\sqrt{\dot{\rho}^2\!+\!1\!-\!
 \frac{8\pi}{3}\varepsilon_{\rm in}\rho^2\!-\!\frac{2m_{\rm in}}{\rho}}
 -\sigma_{\rm out}\sqrt{\dot{\rho}^2\!+\!1\!-\!
 \frac{8\pi}{3}\varepsilon_{\rm out}\rho^2\!-
 \!\frac{2m_{\rm out}}{\rho}}.
 \label{eqmot}
\end{equation}
In this equation $\rho=\rho(\tau)$ is a shell radius, $\tau$ is a
proper time measured by an observer at rest with respect to this
shell, $\dot{\rho}$ is derivative with respect to a proper time,
$\sigma$ is a surface energy density on the shell (in the
discussed vacuum case $\sigma$=const), $m_{\rm in}$ and $m_{\rm
out}$ is the Schwarzschild mass of the inner and outer region
respectively, $\varepsilon_{\rm in}$ is an energy density of the
inner region, $\varepsilon_{\rm out}$ is an energy density of the
outer region and symbols $\sigma_{\rm in,out}=\pm1$. These symbols
equal to $1$, if radius of a two-dimensional sphere is growing in
the direction of an outgoing normal, and equal to $-1$ in the
opposite case.

\begin{figure}[t]
\begin{center}
\includegraphics[width=0.8\textwidth]{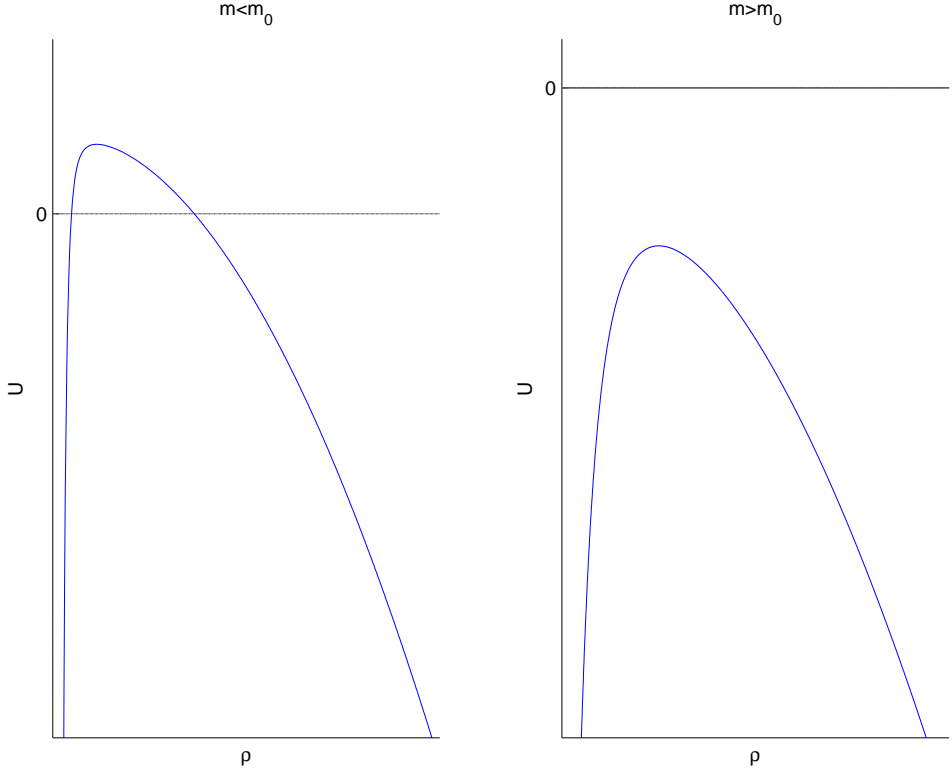}
\end{center}
\caption{\label{figure2} The graphs of potential function
$U(\rho)$ from (\ref{potential}) at  $m<m_0$ (left) and $m>m_0$
(right).}
\end{figure}

For a further analysis we rewrite the equation of motion
(\ref{eqmot}) in the ``energy conservation'' form
$(1/2)\dot{\rho}^2+U(\rho)=0$, with an effective potential
\begin{eqnarray}
 U(\rho) &=&
 \frac{1}{2}-\frac{(m_{\rm out}-m_{\rm in})^2}{32\pi^2\sigma^2\rho^4}-
 \frac{m_{\rm out}+m_{\rm in}}{2\rho}
 -(m_{\rm out}-m_{\rm in})\frac{\varepsilon_{\rm out}-
 \varepsilon_{\rm in}}{12\pi\sigma^2\rho} \nonumber\\
 &&- \frac{(\varepsilon_{\rm in}+\varepsilon_{\rm out}+6\pi\sigma^2)^2-
 4\varepsilon_{\rm in}\varepsilon_{\rm out}}{18\sigma^2}\rho^2,
 \label{potential}
\end{eqnarray}
which is shown in the Fig.~\ref{figure2}. Values of $\sigma_{in}$
and $\sigma_{out}$ in the equation of motion (\ref{eqmot}) in the
``energy conservation'' form are defined by relations
\begin{equation}
 \sigma_{\rm in}={\rm sign}\left[m_{\rm out}-m_{\rm in}
 +\frac{4\pi}{3}(\varepsilon_{\rm out}-\varepsilon_{\rm in})\rho^3
 +8\pi^2\sigma^2\rho^3\right]; \\
 \label{sigmain}
\end{equation}
\begin{equation}
 \sigma_{\rm out}={\rm sign}\left[m_{\rm out}-m_{\rm in}
 +\frac{4\pi}{3}(\varepsilon_{\rm out}-
 \varepsilon_{\rm in})\rho^3-8\pi^2\sigma^2\rho^3\right].
 \label{sigmaout}
\end{equation}
We will consider the general case when a surface energy density of
the shell $\sigma$ may be as positive and negative. It must be
noted that for the positive value of $\sigma$ there is the
exceptional case, when $\sigma_{\rm in}=-1$ and $\sigma_{\rm
out}=1$. In this case the equation of motion (\ref{eqmot}) has no
solution. It would be also seen in the following Carter-Penrose
diagrams.

It is useful to define the following two quantities, one with a
dimension of mass, $m=m_{\rm out}+m_{\rm in}$, and one
dimensionless, $\mu=(m_{\rm out}-m_{\rm in})/(m_{\rm out}+m_{\rm
in})$. With this definitions it is seen from (\ref{sigmain}) that
$\sigma_{\rm in}$ changes its sign at $\rho(\tau)=\rho_{1}$, where
\begin{equation}
 \rho_{1}^3 = \frac{3\mu m}{4\pi(\varepsilon_{\rm in}-
 \varepsilon_{\rm out}-6\pi\sigma^2)},
\end{equation}
provided that (i) $\mu>0$ and $\varepsilon_{\rm
in}>\varepsilon_{\rm out}+6\pi\sigma^2$ or (ii) $\mu<0$ and
$\varepsilon_{\rm in}<\varepsilon_{\rm out}+6\pi\sigma^2$.
Respectively, $\sigma_{\rm out}$ changes its sign when
$\rho(\tau)=\rho_{2}$, where
\begin{equation}
 \rho_{2}^3 =
 \frac{3\mu m}{4\pi(\varepsilon_{\rm in}-\varepsilon_{\rm out}+6\pi\sigma^2)},
\end{equation}
provided, that (iii) $\mu>0$ and $\varepsilon_{\rm
in}>\varepsilon_{\rm out}-6\pi\sigma^2$ or (iv) $\mu<0$ and
$\varepsilon_{\rm in}<\varepsilon_{\rm out}-6\pi\sigma^2$.

Now we consider a behavior of the effective potential
(\ref{potential}). The zeros of this potential, $U(\rho)=0$,
define the bounce points of solution when $\dot\rho=0$. The
maximum of potential (\ref{potential}) is at the point
$\rho=\rho_{\max}$, where
\begin{eqnarray}
 \rho_{\max}^3 &=& my_{\max}=\frac{9\sigma^2}{4}m
 \Bigg\{1+\mu\frac{\varepsilon_{\rm out}-
 \varepsilon_{\rm in}}{6\pi\sigma^2}                 \nonumber\\
 &+& \sqrt{\Big(1\!+\!\mu\frac{\varepsilon_{\rm out}-
 \varepsilon_{\rm in}}{6\pi\sigma^2}\Big)^2\!+
 \!{\frac{2\mu^2}{9\pi^2\sigma^4}
 [(\varepsilon_{\rm in}+\varepsilon_{\rm out}+6\pi\sigma^2)^2-
 4\varepsilon_{\rm in}\varepsilon_{\rm out}]}}\Bigg\} \nonumber\\
 &\Big/&
 \big[(\varepsilon_{\rm in}+\varepsilon_{\rm out}
 +6\pi\sigma^2)^2-4\varepsilon_{\rm in}\varepsilon_{\rm out}\big].
\end{eqnarray}
This maximum corresponds to the zero of potential,
$U(\rho_{\max})=0$, at $m=m_{0}$, where.
\begin{eqnarray}
 m_{0}&=&\sqrt{y_{\max}}\Big/\bigg\{1+\frac{y_{\max}}{9\sigma^2}
 \bigl[(\varepsilon_{\rm in}+\varepsilon_{\rm out}
 +6\pi\sigma^2)^2- 4\varepsilon_{\rm in}\varepsilon_{\rm out}\bigr]
 \nonumber \\
 &&+\mu\frac{\varepsilon_{\rm out}-
 \varepsilon_{\rm in}}{6\pi\sigma^2}+
 \frac{\mu^2}{16\pi^2\sigma^2y_{\max}}\bigg\}^{3/2}.
 \label{m0}
\end{eqnarray}
The potential $U(\rho_{\max})>0$ at $m<m_{0}$ and vise versa. A
second derivative of the potential (\ref{potential}) is
\begin{eqnarray}
 \frac{d^2 U}{d\rho^2} &=&
 -\frac{m}{\rho^3}-\frac{4\pi(\varepsilon_{\rm out}+
 \varepsilon_{\rm in})+12\pi^2\sigma^2}{3} \\
 &&- \frac{9\mu^2m^2}{16\pi^2\sigma^2\rho^6}-
 \left(\frac{\varepsilon_{\rm out}-\varepsilon_{\rm in}}{3\sigma}+
 \frac{\mu m}{4\pi\sigma\rho^3}\right)^2<0. \nonumber
\end{eqnarray}
It is negative everywhere, and so there is no of a stable
equilibrium point for the equation of motion (\ref{eqmot}).

At the next step let us define the values of parameter $m$ when
radii $\rho_{1}$ and $\rho_{2}$ (where $\sigma_{\rm in}$ and
$\sigma_{\rm out}$ change the sign) coincide  with the bounce
points of the equation of motion (\ref{eqmot}). The corresponding
solution of equation $U(\rho_{1})=0$ is $m=m_{1}$, where
\begin{eqnarray}
 m_{1}=
 \sqrt{\frac{3\mu}{4\pi(\varepsilon_{\rm in}-\varepsilon_{\rm out}
 -6\pi\sigma^2)\left(1+\mu\frac{\varepsilon_{\rm in}+
 \varepsilon_{\rm out}+6\pi\sigma^2}{\varepsilon_{\rm in}-
 \varepsilon_{\rm out}-6\pi\sigma^{2}}\right)^{3}}}.
 \label{m1}
\end{eqnarray}
Respectively, the corresponding solution of equation
$U(\rho_{2})=0$ is $m=m_{3}$, where
\begin{eqnarray}
  m_{3}=\sqrt{\frac{3\mu}{4\pi(\varepsilon_{\rm in}-
  \varepsilon_{\rm out}+6\pi\sigma^2)\left(1+
  \mu\frac{\varepsilon_{\rm in}+\varepsilon_{\rm out}
  +6\pi\sigma^2}{\varepsilon_{\rm in}-\varepsilon_{\rm out}
  +6\pi\sigma^{2}}\right)^{3}}}.
 \label{m3}
\end{eqnarray}
By using (\ref{m0}), (\ref{m1}) and (\ref{m3}) it can be shown
that both $m_{1}<m_{0}$ and $m_{3}<m_{0}$.

There are degenerate cases when the inner and outer regions with
respect to the shell have only one event horizon. The inner
Schwarzschild-de Sitter metric has only one event horizon when
$m=m_{21}$, where
\begin{eqnarray}
 m_{21} = \frac{1}{(1-\mu)\sqrt{18\pi\varepsilon_{\rm in}}}.
\end{eqnarray}
Respectively, the outer Schwarzschild-de Sitter metric has only
one event horizon when $m=m_{22}$, where
\begin{eqnarray}
 m_{22} = \frac{1}{(1+\mu)\sqrt{18\pi\varepsilon_{\rm out}}}.
\end{eqnarray}
It can be verified that both $m_{21}>m_{0}$ and $m_{22}>m_{0}$.

For the following analysis of the dynamical evolution of the shell
it is important to know the values of potential (\ref{potential})
at the event horizons of both the inner and outer metrics,
$\rho_{\rm h\,in}=(\rho_{\rm h\,in1},\rho_{\rm h\,in2}, \rho_{\rm
h\,in3})$ and $\rho_{\rm h\,out}=(\rho_{\rm h\,out1}, \rho_{\rm
h\,out2}, \rho_{\rm h\,out3})$, respectively. By using equation
(\ref{horizon}) for the inner event horizon radius, $r=\rho_{\rm
h\,in}$, and equation (\ref{potential}) for potential, after some
algebraic manipulation we obtain
\begin{equation}
 U(\rho_{\rm h\,in})=
 -\frac{\left\{2\left[\frac{\varepsilon_{\rm in}m_{\rm out}}{6\pi\sigma^2}
 \!-\!m_{\rm in}(1\!+\!\frac{\varepsilon_{\rm out}}
 {6\pi\sigma^2})\right]\! + \!\rho_{\rm h\,in}\left(1\!+
 \!\frac{\varepsilon_{\rm out}-
 \varepsilon_{\rm in}}{6\pi\sigma^2}\right)\right\}^2}
  {\left[8\rho_{\rm h\,in}(\rho_{\rm h\,in}-2m_{\rm in})\right]}\leq0.
 \label{u-grin}
\end{equation}
Analogously, for the outer event horizon $r=\rho_{\rm h\,out}$ we
obtain
\begin{equation}
 U(\rho_{\rm h\,out})
 =-\frac{\left\{2\big[\frac{\varepsilon_{\rm out}m_{\rm in}}{6\pi\sigma^2}
 \!-\!m_{\rm out}\left(1\!+\!\frac{\varepsilon_{\rm in}}
 {6\pi\sigma^2}\right)\big]
 +\!\rho_{\rm h\,out}\left(1\!+\!\frac{\varepsilon_{\rm in}-
 \varepsilon_{\rm out}}{6\pi\sigma^2}\right)\right\}^2}
 {\left[8\rho_{\rm h\,out}(\rho_{\rm h\,out}-2m_{\rm out})\right]}\leq0.
 \label{u-grout}
\end{equation}
The equality in (\ref{u-grin}) and (\ref{u-grout}) is achieved
only at $m=m_{1}$ (inner metric), where $\rho_{\rm
h\,in}=\rho_{1}$, and, respectively, at $m=m_{3}$ (outer metric),
where $\rho_{\rm h\,out}=\rho_{2}$. In other words, the points of
event horizons of both the inner and outer metrics, $\rho_{\rm
h\,in}$ and $\rho_{\rm h\,out}$ cannot be below the potential in
the Fig.~\ref{figure2}. The gravitational radius $\rho_{\rm h2}$
for both `in' and `out' metrics is changed in the range
$0\leq\rho_{\rm h2}\leq\rho_{\rm h1}$, and, therefore, it is
placed at the left of the potential curve. In a similar way, the
gravitational radius $\rho_{\rm h3}$ is changed in the range
$\rho_{\rm h1}\leq\rho_{\rm h3}\leq\sqrt{3}\rho_{\rm h1}$, and,
therefore, it is placed at the right of the potential curve.

Now we have all necessary ingredients for the investi\-gation of
possible motions of the thin vacuum shell in the Schwarzschild-de
Sitter metric.

\section{Dynamical evolution of vacuum shell}
\label{sec3}

\subsection{Case of $\mu>0$ and $\varepsilon_{\rm out}>\varepsilon_{\rm
 in}+6\pi\sigma^2$}
 \label{sec3.1}

\begin{figure}[t]
\begin{center}
\includegraphics[width=0.8\textwidth]{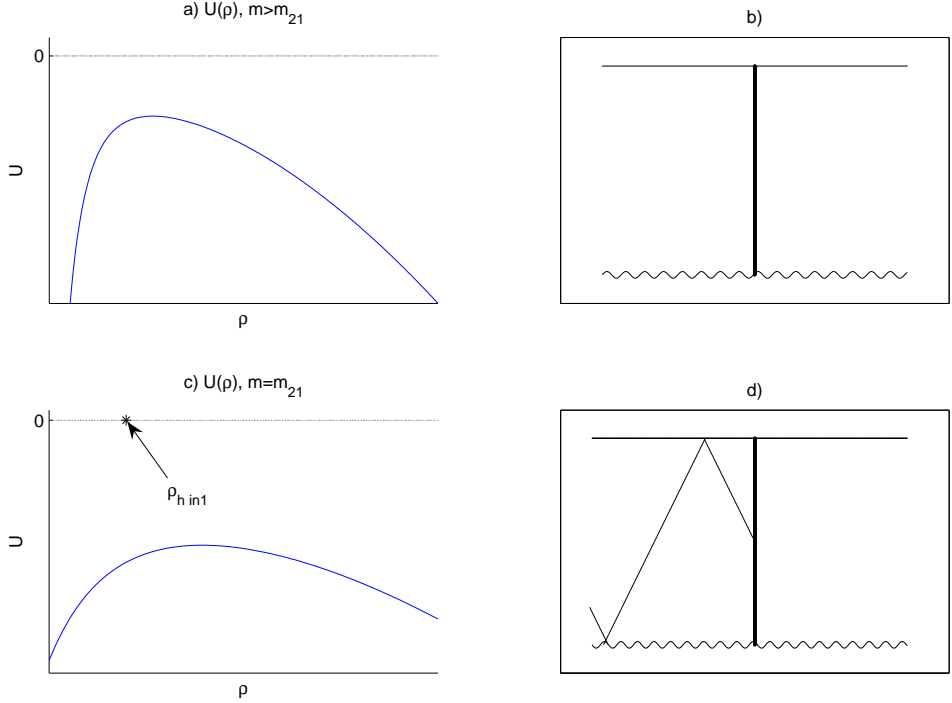}
\end{center}
\caption{\label{figure3} An effective potential (\ref{potential})
and the corresponding Carter-Penrose diagrams for the case
$\mu>0$, $\varepsilon_{\rm out}>\varepsilon_{\rm in}+6\pi\sigma^2$
and values of the rolling mass parameter $m$ in the range
$m>m_{21}$ (graphs a and b) and $m=m_{21}$ (graphs c and d). See
Section~3.1 for details.}
\end{figure}

First of all we consider a simple case, when $\mu>0$ and
$\varepsilon_{\rm out}>\varepsilon_{\rm in}+6\pi\sigma^2$ (another
similar case when $\mu<0$ and $\varepsilon_{\rm
in}>\varepsilon_{\rm out}+6\pi\sigma^2$ may be analyzed in a
similar way). In this case $\sigma_{\rm in}=\sigma_{\rm out}=1$,
as it follows from (\ref{sigmain}) and (\ref{sigmaout}).
Therefore, there are no radii $\rho_{1}$ and $\rho_{2}$. It can be
shown that in this case $m_{21}>m_{22}$, $\rho_{\rm
h\,in1}>\rho_{\rm h\,out1}$ and $\rho_{\rm h\,in3}>\rho_{\rm
h\,out3}$.

The rolling parameter of our classification scheme is a mass
parameter $m$ and we start from the large value of this parameter.

If $m>m_{21}$, then an event horizon is absent and initial
expansion (or contraction) of the shell is unbounded (there is no
bounce point). The corresponding potential $U(\rho)$ and the
Carter-Penrose diagram is shown in the Figs.~\ref{figure3}a and
\ref{figure3}b respectively. Here and further below the
Carter-Penrose diagrams are shown only for an initially expanding
envelope. The corresponding diagrams for contracting envelope are
easily reproduced from the expanding ones by symmetry reflection
with respect to the median horizontal line (except for some
special cases).

At $m=m_{21}$, the first event horizon appears in the inner
metric, $\rho_{\rm h\,in1}$. See Figs.~\ref{figure3}c and
\ref{figure3}d for the corresponding potential and diagram.

\begin{figure}[t]
\begin{center}
\includegraphics[width=0.8\textwidth]{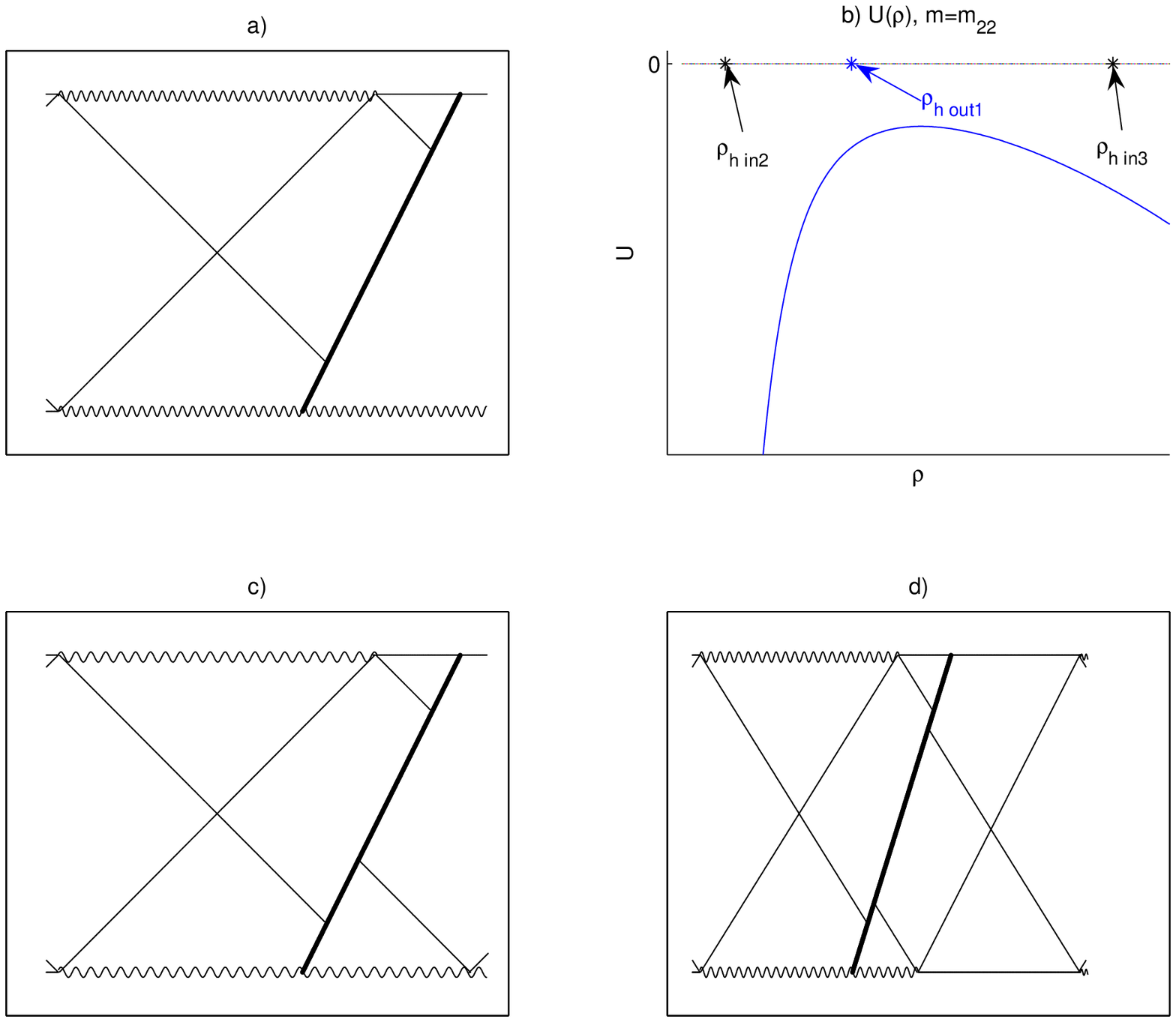}
\end{center}
\caption{\label{figure4} An effective potential (\ref{potential})
and the corresponding Carter-Penrose diagrams for the case
$\mu>0$, $\varepsilon_{\rm out}>\varepsilon_{\rm in}+6\pi\sigma^2$
and values of the rolling mass parameter $m$ in the range
$m_{22}<m<m_{21}$ (graph a), $m=m_{22}$ (graphs b and c) and
$m_{22}>m>\max(m_{0},m^{\ast})$ (graph d).}
\end{figure}

If $m_{22}<m<m_{21}$, then there are two event horizons,
$\rho_{\rm h\,in2}$ and $\rho_{\rm h\,in3}$. See
Fig.~\ref{figure4}a for the Carter-Penrose diagram, while the
potential has a similar form as in the previous two cases
(Figs.~\ref{figure4}a and \ref{figure3}c).

At $m=m_{22}$, the first event horizon appears in the outer
metric, $\rho_{\rm h\,out1}<\rho_{\rm h\,in3}$. We prove that
$\rho_{\rm h\,in2}<\rho_{\rm h\,out1}$. Indeed, this inequality
may be written in the form $(1+\mu)/(1-\mu)>2\varepsilon_{\rm
out}/(3\varepsilon_{\rm out}-\varepsilon_{\rm in})$. For $\mu>0$,
if we will prove the inequality $1>2\varepsilon_{\rm
out}/(3\varepsilon_{\rm out}-\varepsilon_{\rm in})$, then we prove
our statement. The last inequality is evident by remembering that
$\varepsilon_{\rm out}>\varepsilon_{\rm in}+6\pi\sigma^2$. So for
$m=m_{22}$ the event horizon $\rho_{\rm h\,out1}$ is located
between $\rho_{\rm h\,in2}$ and $\rho_{\rm h\,in3}$. See
Figs.~{\ref{figure4}}b and {\ref{figure4}}c for the corresponding
potential and the Carter-Penrose diagram of an expanding shell.

If $m_{22}>m>\max(m_{0},m^{\ast})$, where $m^{\ast}$ is defined
from equation $\rho_{\rm h\,in2}=\rho_{\rm h\,out2}$, there are
two event horizons, $\rho_{\rm h\,out2}$ and $\rho_{\rm h\,out3}$,
in the outer metric. These event horizons are located between the
event horizons of the inner metric $\rho_{\rm h\,in2}$ and
$\rho_{\rm h\,in3}$. See Fig.~\ref{figure4}d for a corresponding
diagram.

If $m^{\ast}>m_{0}$ and $m^{\ast}>m>m_{0}$, then the arrangement
of event horizons is $\rho_{\rm h\,out2}<\rho_{\rm
h\,in2}<\rho_{\rm h\,out3}<\rho_{\rm h\,in3}$. The Carter-Penrose
diagram for an expanding shell is shown in the
Fig.~\ref{figure8}d.

\begin{figure}[t]
\begin{center}
\includegraphics[width=0.8\textwidth]{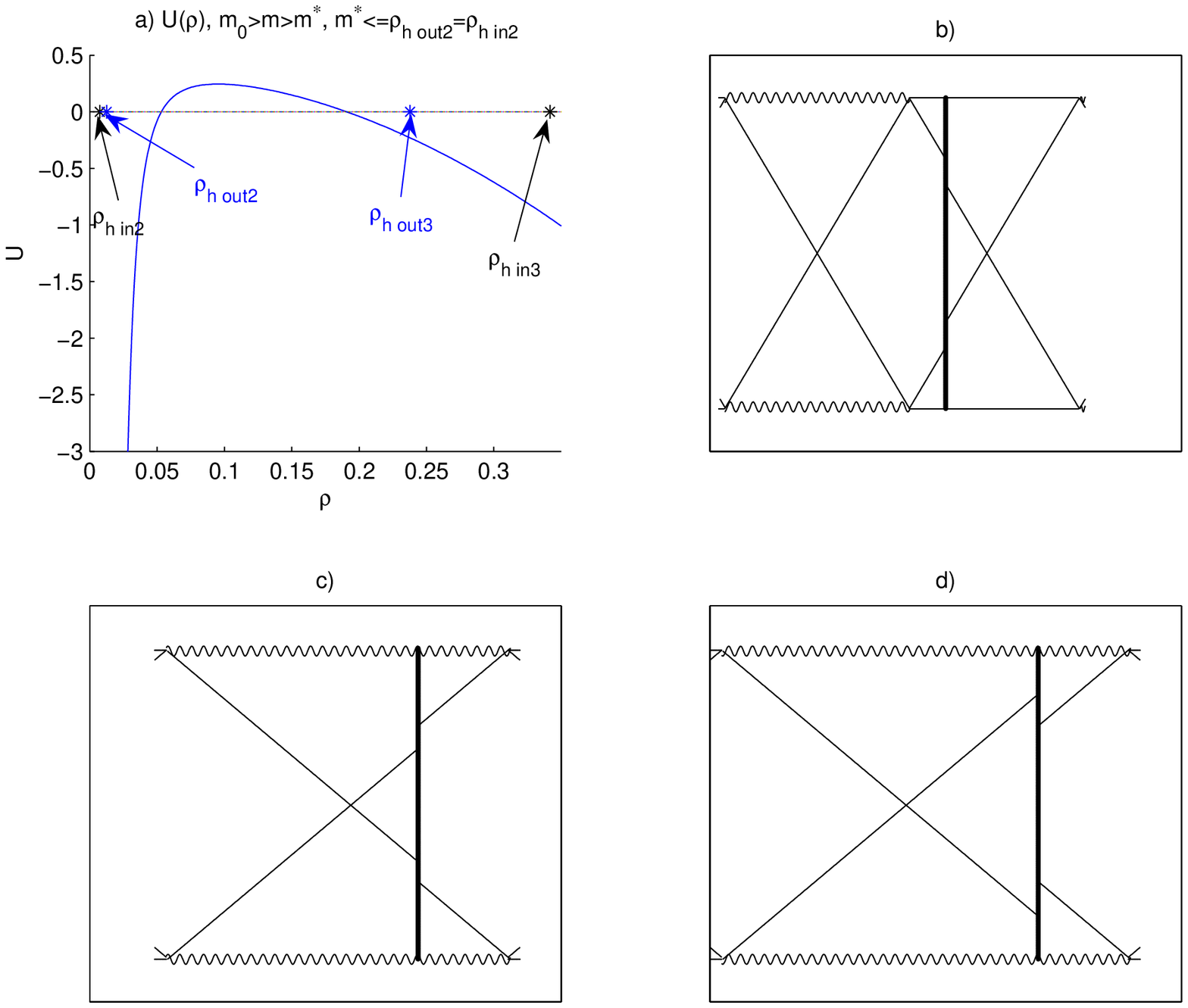}
\end{center}
\caption{\label{figure5} An effective potential (\ref{potential})
and the corresponding Carter-Penrose diagrams for the case
$\mu>0$, $\varepsilon_{\rm out}>\varepsilon_{\rm in}+6\pi\sigma^2$
and values of the rolling mass parameter $m$ in the range
$m<m_{0}$ (graphs a, b and c), $m^{\ast}<m<m_{0}$ (graphs b and d)
and $m<m^{\ast}<m_{0}$ (graphs b and c).}
\end{figure}

If $m<m_{0}$, the region appears, where potential $U(\rho)$ is
positive (see Fig.~\ref{figure5}a). Now solutions for $\rho(\tau)$
have the bounce points. The arrangement of event horizons is
$\rho_{\rm h\,out2}<\rho_{\rm h\,in2}<\rho_{\rm h\,out3}<\rho_{\rm
h\,in3}$. The Carter-Penrose diagrams for an expanding and
contracting shell are shown in the Fig.~\ref{figure5}b and
Fig.~\ref{figure5}c respectively.

If $m_{0}>m^{\ast}$, then in the case $m^{\ast}<m<m_{0}$ the
possible Carter-Penrose diagrams for an expanding and contracting
shell are shown in the Figs.~\ref{figure5}b and \ref{figure5}d.

In the last case $m<m^{\ast}$ the arrangement of event horizons is
$\rho_{\rm h\,out2}<\rho_{\rm h\,in2}<\rho_{\rm h\,out3}<\rho_{\rm
h\,in3}$. See in the Figs.~\ref{figure5}b and \ref{figure5}c the
Carter-Penrose  diagrams for an expanding and contracting shell
respectively. In particular, these diagrams illustrate the
formation of a black hole or wormhole after a final contraction of
the shell.

\subsection{Case of $\mu>0$ and $\varepsilon_{\rm in}>
\varepsilon_{\rm out}+6\pi\sigma^2$}

\begin{figure}[t]
\begin{center}
\includegraphics[width=0.8\textwidth]{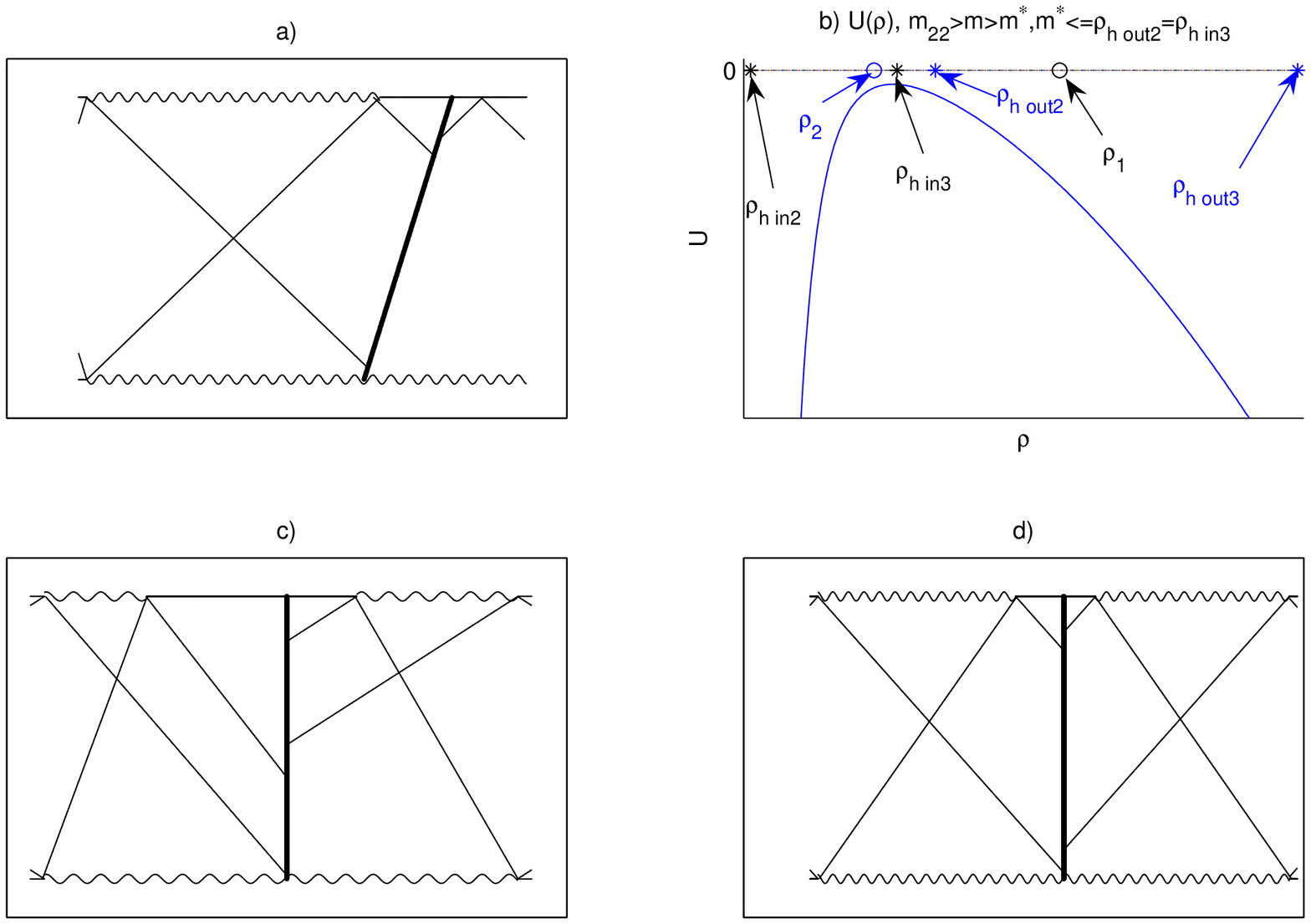}
\end{center}
\caption{\label{figure6}  An effective potential (\ref{potential})
and the corresponding Carter-Penrose diagrams for the case $\mu>0$
and $\varepsilon_{\rm in}> \varepsilon_{\rm out}+6\pi\sigma^2$ and
values of the rolling mass parameter $m$ in the range $m=m_{22}$
(graphs a), $m_{22}>m>m^{\ast}$ (graphs b and c) and
$m^{\ast}>m>m_{0}$ (graph d).}
\end{figure}

From this point we start classification of a more tangled case,
when $\mu>0$ and $\varepsilon_{\rm in}>\varepsilon_{\rm
out}+6\pi\sigma^2$ (the case when $\mu<0$ and $\varepsilon_{\rm
out}>\varepsilon_{\rm in}+6\pi\sigma^2$ may be analyzed in a
similar way). Now both $\sigma_{\rm in}$ and $\sigma_{\rm out}$
can change the signs at radii $\rho_{1}$ and $\rho_{2}$. In the
considered case the following inequalities are valid:
$\rho_{1}>\rho_{2}$ and $\rho_{\rm h\,out1}>\rho_{\rm h\,in1}$.

It is useful to define the additional 5 dimensionless parameters:
$\tilde{\mu}$, $\mu_{\rm out}$, $\mu_{\rm in}$, $\mu_{\rm 1\,out}$
and $\mu_{\rm 1\,in}$. A first parameter is a solution
$\mu=\tilde{\mu}$ of the equation $m_{21}=m_{22}$, where
\begin{equation}
 \tilde{\mu} =
 \frac{\sqrt{\varepsilon_{\rm in}}-\sqrt{\varepsilon_{\rm out}}}
 {\sqrt{\varepsilon_{\rm in}}+\sqrt{\varepsilon_{\rm out}}}.
\end{equation}
($m_{22}<m_{21}$ at $\mu>\tilde{\mu}$ and vice versa). A second
parameter is a solution $\mu=\mu_{\rm out}$ of the equation
$\rho_{2}=\rho_{\rm h\,out1}$ at $m=m_{22}$, where
\begin{equation}
 \mu_{\rm out}=
 \frac{\varepsilon_{\rm in}-\varepsilon_{\rm out}+6\pi\sigma^2}
 {5\varepsilon_{\rm out}-\varepsilon_{\rm in}-6\pi\sigma^2}.
\end{equation}
This parameter exists only if $3\varepsilon_{\rm
out}>\varepsilon_{\rm in}+6\pi\sigma^2$. Under this condition
$\rho_{2}>\rho_{\rm h\,out1}$ at $\mu>\mu_{\rm out}$ and vice
versa. In the opposite case, when $3\varepsilon_{\rm
out}<\varepsilon_{\rm in}+6\pi\sigma^2$, it is always
$\rho_{2}<\rho_{\rm h\,out1}$. A third parameter is a solution
$\mu=\mu_{\rm in}$ of the equation $\rho_{1}=\rho_{\rm h\,in1}$ at
$m=m_{21}$, where
\begin{equation}
 \mu_{\rm in} =
 \frac{\varepsilon_{\rm in}-\varepsilon_{\rm out}-6\pi\sigma^2}
 {5\varepsilon_{\rm in}-\varepsilon_{\rm out}-6\pi\sigma^2}
\end{equation}
($\rho_{1}>\rho_{\rm h\,in1}$ at $\mu>\mu_{\rm in}$ and vice
versa). A fourth parameter is a solution $\mu=\mu_{\rm 1\,out}$ of
the equation $\rho_{\rm h\,out1}=\rho_{\rm h\,in3}$ at $m=m_{22}$,
where
\begin{equation}
 \mu_{\rm 1\,out}=\frac{\varepsilon_{\rm in}-\varepsilon_{\rm out}}
 {5\varepsilon_{\rm out}-\varepsilon_{\rm in}},
\end{equation}
This parameter exists only if $\varepsilon_{\rm
in}<3\varepsilon_{\rm out}$. Under this condition $\rho_{\rm
h\,out1}<\rho_{\rm h\,in3}$ at $\mu>\mu_{1\,out}$ and vise versa.
In the opposite case, when $\varepsilon_{\rm in}>3\varepsilon_{\rm
out}$, it is always $\rho_{\rm h\,out1}>\rho_{\rm h\,in3}$.
Finally, the fifth parameter is a solution $\mu=\mu_{\rm 1\,in}$
of the equation $\rho_{\rm h\,in1}=\rho_{\rm h\,out2}$ at
$m=m_{21}$, where
\begin{equation}
 \mu_{\rm 1\,in}=
 \frac{\varepsilon_{\rm in}-\varepsilon_{\rm out}}
 {5\varepsilon_{\rm in}-\varepsilon_{\rm out}}
\end{equation}
($\rho_{\rm h\,in1}>\rho_{\rm h\,out2}$ at $\mu<\mu_{1\,in}$ and
vice versa). It is easy to verify that $\rho_{\rm
h\,out1}>\rho_{\rm h\,in2}$ at $m=m_{22}$ and $\rho_{\rm
h\,in1}<\rho_{\rm h\,out3}$ at $m=m_{21}$. A mutual arrangement of
these 5 parameters is $\mu_{\rm
out}>\mu_{1\,out}>\tilde{\mu}>\mu_{1\,in}>\mu_{\rm in}$.

As a first step in classification of the situation, when $\mu>0$
and $\varepsilon_{\rm in}>\varepsilon_{\rm out}+6\pi\sigma^2$, we
consider the case, when $\mu>\tilde{\mu}$ (the case
$\mu<\tilde{\mu}$ is quite a similar). Again we begin from the
large value of the rolling parameter $m$.

If $m>m_{21}$, the event horizons are absent and there are only
two radii $\rho_{1}$ and $\rho_{2}$, where the signs of
$\sigma_{in,out}$ are changed. The potential $U(\rho)$ is similar
to one shown in the Fig.~\ref{figure3}a (where $\rho_{1}$ and
$\rho_{2}$ are not shown). A diagram for an expanding shell is
analogous to one shown in the Fig.~\ref{figure3}b.

The next case is $m=m_{21}$. Now a first event horizon $\rho_{\rm
h\,in1}$ appears, which is at left to $\rho_{1}$ (i.~e. $\rho_{\rm
h\,in1}<\rho_{1}$). Now potential is similar to one in the
Fig.~\ref{figure3}c. The crucial point is that now a shell
intersects the radius $\rho_{\rm h\,in1}$, when $\sigma_{\rm
in}=1$. The Carter-Penrose diagram is similar to one shown in the
Fig.~\ref{figure3}d.

If $m_{21}>m>m_{22}$, there are two event horizons $\rho_{\rm
h\,in2}$ and $\rho_{\rm h\,in3}$, but qualitatively this case is
similar to the preceding one. The Carter-Penrose diagram for an
expanding shell is shown in the Fig.~\ref{figure4}a.

The case $m=m_{22}$ is divided into subcases. We begin from the
subcase $\varepsilon_{\rm in}>3\varepsilon_{\rm out}$, when
inequalities $\rho_{\rm h\,out1}>\rho_{\rm h\,in3}$ and $\rho_{\rm
h\,out1}>\rho_{2}$ are fulfilled. The resulting diagram for an
expanding shell is shown in the Fig.~\ref{figure6}a. The other
subcase will be considered later.

If $m_{22}>m>m^{\ast}$, where $m^{\ast}$ is a solution of the
equation $\rho_{\rm h\,out2}=\rho_{\rm h\,in3}$. Now instead of
one event horizon $\rho_{\rm h\,out1}$ there are two event
horizons, $\rho_{\rm h\,out2}$ and $\rho_{\rm h\,out3}$. The
corresponding potential and the Carter-Penrose diagram are shown
in the Fig.~\ref{figure6}b and \ref{figure6}c respectively. It
must be noted that a moving shell intersects the event horizons
$\rho_{\rm h\,out}$ when $\sigma_{\rm out}=-1$. Therefore, a shell
intersects these event horizons in the region $R_{-}$. On the
contrary, a shell intersects the event horizons $\rho_{\rm h\,in}$
in the region $R_{+}$ when $\sigma_{\rm in}=1$.

The only distinguishing feature of the case $m^{\ast}>m>m_{0}$
from the preceding one is swapping round the event horizons
$\rho_{\rm h\,out2}$ and $\rho_{\rm h\,in3}$. The arrangement of
radii is ($\rho_{\rm h\,in2},\rho_{2})<\rho_{\rm
h\,out2}<\rho_{\rm h\,in3}<(\rho_{1},\rho_{\rm h\,out3})$. A
corresponding diagram is shown in the Fig.~\ref{figure6}d.

\begin{figure}[t]
\begin{center}
\includegraphics[width=0.8\textwidth]{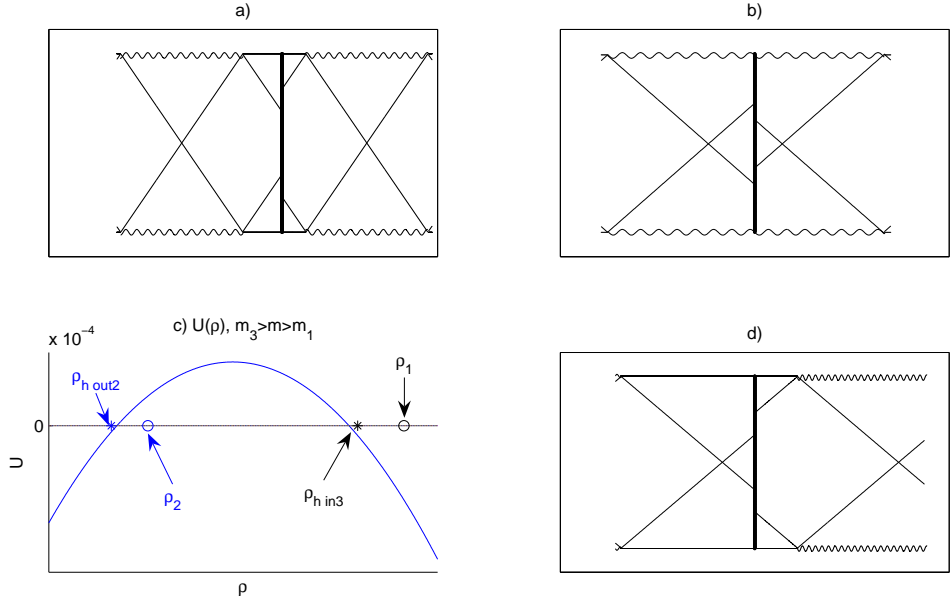}
\end{center}
\caption{\label{figure7} An effective potential (\ref{potential})
and the corresponding Carter-Penrose diagrams for the case
$\mu>0$, $\varepsilon_{\rm out}>\varepsilon_{\rm in}+6\pi\sigma^2$
and values of the rolling mass parameter $m$ in the range
$m_{0}>m>m_{3}$ (graphs a and b), $m_{3}>m>m_{1}$ (graphs c) and
$m<m_{1}$ (graph d).}
\end{figure}

If $m_{0}>m>m_{3}$, the potential intersects the axis $U=0$ and a
shell will bounce from the potential. The arrangements of radii is
the same as in the previous case. The diagrams for expanding and
contracting shell are shown in the Fig.~\ref{figure7}a and
\ref{figure7}b respectively.

If $m_{3}>m>m_{1}$, the radii $\rho_{\rm h\,out2}$ and $\rho_{2}$
are swapped round. The potential is shown in the
Fig.~\ref{figure7}c (the event horizons $\rho_{\rm h\,out3}$ and
$\rho_{\rm h\,in2}$ are not shown). A diagram for an expanding
shell is the same as in the Fig.~\ref{figure7}a. A contracting
shell intersects the event horizon $\rho_{\rm h\,out2}$ in the
region $R_{+}$ (when $\sigma_{\rm out}=1$). A diagram for a
contracting shell is shown in the Fig.~\ref{figure5}d.

If $m<m_{1}$, the radii $\rho_{\rm h\,in3}$ and $\rho_{1}$ are
swapped round. Two radii $\rho(1)$ and $\rho(2)$ (where signs of
$\sigma_{in,out}$ are changed) is now under the potential graph.
The diagram for a contracting shell is the same as in the previous
case (see Fig.~\ref{figure5}d). A corresponding diagram for an
expanding shell is shown in the Fig.~\ref{figure7}d.

\begin{figure}[t]
\begin{center}
\includegraphics[width=0.8\textwidth]{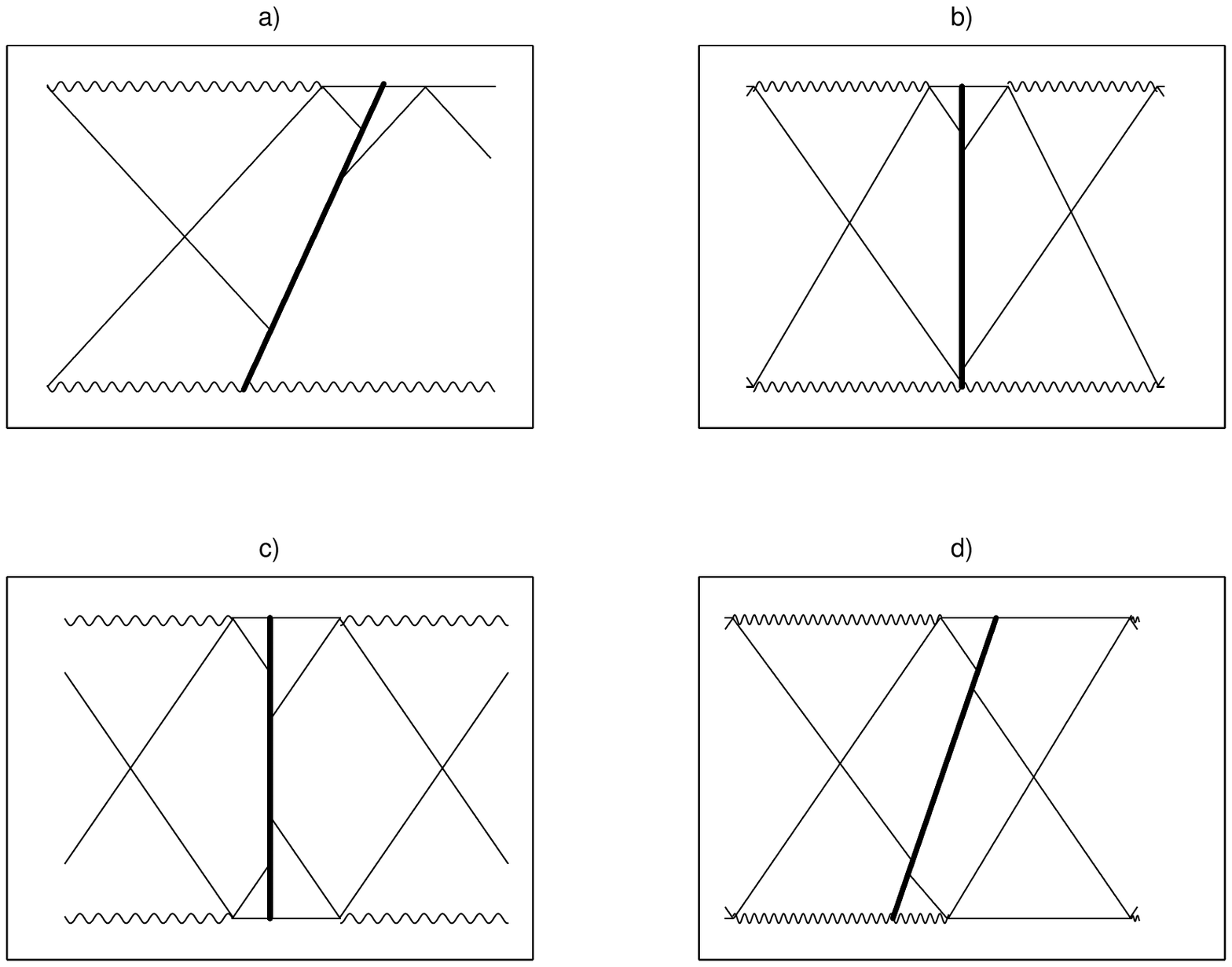}
\end{center}
\caption{\label{figure8} An effective potential (\ref{potential})
and the corresponding Carter-Penrose diagrams for the case
$\mu>0$, $\varepsilon_{\rm out}>\varepsilon_{\rm in}+6\pi\sigma^2$
and values of the rolling mass parameter $m$ in the range
$\mu>\mu_{1\,out}$ and $\mu>\mu_{1\,out}$ (graphs a),
$m_{22}>m>\max(m_{0},m^{\ast})$ (graphs b) and $m_{0}>m>m_{3}$
(graphs c).}
\end{figure}

Now we return to another subcase of the case $m=m_{22}$, when
$\varepsilon_{\rm in}+6\pi\sigma^2>3\varepsilon_{\rm
out}>\varepsilon_{\rm in}$, the inequality $\rho_{2}<\rho_{\rm
h\,out1}$ is fulfilled and parameter $\mu_{1\,out}$ exists. If
$\mu>\mu_{1\,out}$, then $\rho_{\rm h\,out1}<\rho_{\rm h\,in3}$.
The arrangement of radii is ($\rho_{\rm
h\,in2},\rho_{2})<\rho_{\rm h\,out1}<\rho_{\rm h\,in3}<\rho_{1}$.
A diagram for an expanding shell is shown in the
Fig.~\ref{figure8}a.

If $m_{22}>m>\max(m_{0},m^{\ast})$, where $m^{\ast}$ defined from
equation $\rho_{\rm h\,in3}=\rho_{\rm h\,out3}$, instead of the
one event horizon $\rho_{\rm h\,out1}$ there are two event
horizons $\rho_{\rm h\,out2,3}$. The arrangement of radii is
$(\rho_{\rm h\,in2},\rho_{2})<\rho_{\rm h\,out2}<\rho_{\rm
h\,out3}<\rho_{\rm h\,in3}<\rho_{1}$. A corresponding diagram for
an expanding shell is shown in the Fig.~\ref{figure8}b.

If $m^{\ast}>m>m_{0}$, the arrangement of radii is $(\rho_{\rm
h\,in2},\rho_{2})<\rho_{\rm h\,out2}<\rho_{\rm
h\,in3}<(\rho_{1},\rho_{\rm h\,out3}$). The only difference with
preceding subcase is swapping round the event horizons $\rho_{\rm
h\,in3}$ and $\rho_{\rm h\,out3}$. A diagram for an expanding
shell shell shown in the Fig.~\ref{figure6}d.

If $m_{0}>m>m_{3}$, the potential intersects the axis $U=0$ and
the bounce point appears. The arrangement radii is a similar to
the previous case, but a shell now can bounce from the potential.
The diagrams for an expanding and contracting shells are shown in
the Figs.~\ref{figure7}a and the \ref{figure7}b respectively. At
$m_{0}>m^{\ast}$ and at $m_{0}>\max(m_{3},m^{\ast})$ there will be
the following arrangement of radii : $(\rho_{\rm
h\,in2},\rho_{2})<\rho_{\rm h\,out2}<\rho_{\rm h\,out3}<\rho_{\rm
h\,in3}<\rho_{1}$. The diagrams for an expanding and contra\-cting
shells are shown in the Figs.~\ref{figure8}c and \ref{figure7}b
respectively. The next two subcases depend on the relation
$m_{3}\lessgtr m^{\ast}$ and are described in a similar way. The
remaining cases for $m<(m_{3},m^{\ast})$ are similar to ones for
$\varepsilon_{\rm in}>3\varepsilon_{\rm out}$.

If $\mu_{1\,out}>\mu>\tilde{\mu}$ in the case $\varepsilon_{\rm
in}+6\pi\sigma^2>3\varepsilon_{\rm out}>\varepsilon_{\rm in}$,
then $\rho_{\rm h\,out1}>\rho_{\rm h\,in3}$. As a result this case
is reduced to the case, when $\varepsilon_{\rm
in}>3\varepsilon_{\rm out}$.

If $3\varepsilon_{\rm out}-6\pi\sigma^2>\varepsilon_{\rm in}$ for
$m=m_{22}$, then the inequality $\rho_{\rm h\,out2}>\rho_{\rm
h\,in2}$ is fulfilled and besides the parameter $\mu_{1\,out}$
there exists also $\mu_{\rm out}$. At $\mu>\mu_{\rm out}$, the
inequalities $\rho_{\rm h\,out1}<\rho_{\rm h\,in3,}$ and
$\rho_{2}>\rho_{\rm h\,out1}$ are valid. As a result the
arrangement of radii is $\rho_{\rm h\,in2}<\rho_{\rm
h\,out1}<(\rho_{2},\rho_{\rm h\,in3})<\rho_{1}.$ A corresponding
diagram for an expanding shell is shown in the
Fig.~\ref{figure4}c.

If $m_{22}>m>m_{0}$, instead of the one event horizon $\rho_{\rm
h\,out1}$ there are two event horizons $\rho_{\rm h\,out2}$ and
$\rho_{\rm h\,out3}$ and the arrangement of radii is $\rho_{\rm
h\,in2}<\rho_{\rm h\,out2}<\rho_{\rm h\,out3}<(\rho_{2},\rho_{\rm
h\,in3})<\rho_{1}$. A corresponding diagram for an expanding shell
is shown in the Fig.~\ref{figure4}d.

If $m_{0}>m>m_{3}$, the potential intersects the axis $U=0$ and a
shell can bounce from the potential. The corresponding diagrams
for an expanding and contracting shells are shown in the
Figs.~\ref{figure5}b and \ref{figure5}d respectively.

If $m_{3}>m>m^{\ast}$, where $m^{\ast}$ is a solution of the
equation $\rho_{\rm h\,out3}=\rho_{\rm h\,in3}$, the event horizon
$\rho_{\rm h\,out3}$ changes the place with $\rho_{2}$. The
coincidence of these radii occurs at the radius, where
$U(\rho)=0$. The radius $\rho_{2}$ now is under the potential
curve. A shell will intersect $\rho_{\rm h\,out3}$ with
$\sigma_{\rm out}=-1$. A corresponding diagram for an expanding
shell is shown in the Fig.~\ref{figure8}c.

The case $m^{\ast}>m$ is similar to one for $\varepsilon_{\rm
in}>3\varepsilon_{\rm out}$. The case $\mu_{\rm
out}>\mu>\mu_{1\,out}$ is similar to one for $\varepsilon_{\rm
in}+6\pi\sigma^2>3\varepsilon_{\rm out}>\varepsilon_{\rm in}$.
Finally, the case $\mu_{1\,out}>\mu>\tilde{\mu}$ is similar to one
for $\varepsilon_{\rm in}>3\varepsilon_{\rm out}$. Analogously is
considered the case, when $\tilde{\mu}>\mu>0$, and the case, when
exists only one of the two radii, $\rho_{1}$ or $\rho_{2}$. This
case completes the classification.

\section{Conclusion}
\label{sec4}

We classified all possible evolution scenarios of a thin vacuum
shell in the Schwarzschild-de\,Sitter metric and constructed the
Carter-Penrose diagrams for corres\-ponding global geometries.
These geometries illustrates the possibilities for final formation
of black holes and wormholes or eventual expansion of bubbles.

\ack

We acknowledge Viktor Berezin and Yury Erochenko for helpful
discussions. This work was supported in part by the Russian
Foundation for Basic Research grants 06-02-16342-a and the Russian
Ministry of Science grants LSS 5573.2006.2.

\end{document}